# Nonreciprocal thermal radiation in ultrathin magnetized epsilon-near-zero semiconductors

Mengqi Liu[1,2#], Shuang Xia[3,4#], Wenjian Wan[5#], Jun Qin[3,4], Hua Li[5,6]*, Changying Zhao[1]*, Lei Bi[3,4]*, Cheng-Wei Qiu[2]*

[1]Institute of Engineering Thermophysics, MOE Key Laboratory for Power Machinery and Engineering, School of Mechanical Engineering, Shanghai Jiao Tong University, Shanghai 200240, China

[2]Department of Electrical and Computer Engineering, National University of Singapore, Singapore 117583, Singapore

[3]National Engineering Research Center of Electromagnetic Radiation Control Materials, University of Electronic Science and Technology of China, Chengdu 610054, China

[4]State Key Laboratory of Electronic Thin-Films and Integrated Devices, University of Electronic Science and Technology of China, Chengdu, 610054, China

[5]Key Laboratory of Terahertz Solid State Technology, Shanghai Institute of Microsystem and Information Technology, Chinese Academy of Sciences, Shanghai, 200050 China

[6]Center of Materials Science and Optoelectronics Engineering, University of Chinese Academy of Science, Beijing 1000049, China

\* Email: hua.li@mail.sim.ac.cn; changying.zhao@sjtu.edu.cn; bilei@uestc.edu.cn; chengwei.qiu@nus.edu.sg

[#] These authors contributed equally to this work.

## Abstract

Spectral/angular emissivity $e$ and absorptivity $\alpha$ of an object are widely believed to be identical by Kirchhoff's law of thermal radiation in reciprocal systems, but this introduces an intrinsic and inevitable energy loss for energy conversion and harvesting devices. So far, experimental evidences of breaking this well-known balance are still absent, and previous theoretical proposals are restricted to narrow single-band nonreciprocal radiation. Here we observe for the first time, to our knowledge, the violation of Kirchhoff's law using ultrathin ($<\lambda/40$, $\lambda$ is the working wavelength) magnetized InAs semiconductor films at epsilon-near-zero (ENZ) frequencies. Large difference of $|\alpha\text{-}e|>0.6$ has been experimentally demonstrated under a moderate external magnetic field. Moreover, based on magnetized ENZ building blocks supporting asymmetrically radiative Berreman and surface ENZ modes, we show versatile shaping of nonreciprocal thermal radiation: single-band, dual-band, and broadband nonreciprocal emission spectra at different wavebands. Our findings of breaking Kirchhoff's law will advance the conventional understanding of emission and absorption processes of natural objects, and lay a solid foundation for more comprehensive studies in designing various nonreciprocal thermal emitters. The reported recipe of diversely shaping nonreciprocal emission will also breed new possibilities in renovating next-generation nonreciprocal energy devices in the areas of solar cells, thermophotovoltaic, radiative cooling, etc.

Any object with a non-zero temperature would emit and absorb thermal radiation. The object's emissivity $e(\omega,\theta)$ and absorptivity $\alpha(\omega,\theta)$, with $\omega$ and $\theta$ being angular frequency and incident angle respectively, are widely perceived equal according to Kirchhoff's law[1] of thermal radiation:

$$e(\omega,\theta) = \alpha(\omega,\theta). \tag{1}$$

Over the last centuries, this classic law has been extensively deployed by default from macro- to nano-scale thermal radiation[2–5] and served as a guideline to design thermal emitters which play essential roles in a plethora of applications such as solar photovoltaic[6], radiative cooling[7–9], thermal camouflage[10,11], far/near-field thermophotovoltaic[12], gas sensing[13], etc. But this law introduces an intrinsic and inevitable loss for energy conversion and harvesting processes. For example, as for solar absorbers, the ideal situation is to absorb all the incoming solar energy [$\alpha(\omega,\theta) \to 1$] but suppress emission from themselves [$e(\omega,\theta) \to 0$]. In contrast, in terms of radiative cooling devices[7], one may expect to emit thermal radiation to the external environment within atmospheric windows as much as possible [$e(\omega,\theta) \to 1$], with no absorbed energy from the background [$\alpha(\omega,\theta) \to 0$]. Such practical demands motivate us to break Kirchhoff's law and maximize the difference between emissivity and absorptivity: $|\alpha(\omega,\theta) - e(\omega,\theta)| \to 1$. In principle, Kirchhoff's law is originated from Lorentz reciprocity of Maxwell's equations[14], not constrained by the thermodynamic laws. It implies that the balance stated in **Eq. (1)** can be violated by designing nonreciprocal thermal emitters with broken electromagnetic reciprocity. A recent theoretical work has also proven that the nonreciprocal multi-junction solar cells[15], where each layer presents nonreciprocal semi-transparent absorption/emission properties and varying bandgaps, are capable of reaching the Landsberg limit (93.3%) ideally. Therefore, the violation of Kirchhoff's law of thermal radiation is of fundamental and practical significance in both basic theory and energy-based applications.

Though nonreciprocal thermal radiation is highly demanded for next-generation energy devices, it is very challenging to realize significant violation of Kirchhoff's law in experiments. Without losing generality, we begin with revisiting the thermodynamic constraints[14] on an opaque thermal emitter as sketched in **Fig. 1a** that undergoes energy exchanges with two radiative channels denoted with blackbodies 1 and 2 under thermal equilibrium. According to the second thermodynamic law, there are $\alpha_1 + R_{1\to 2} = 1$, $e_1 + R_{2\to 1} = 1$ and $\alpha_2 + R_{2\to 1} = 1$, $e_2 + R_{1\to 2} = 1$, regardless of the thermal emitter being nonreciprocal or not. The $R_{2\to 1}$ and $R_{1\to 2}$ represent the specular reflectivity. Thus, one can directly obtain the relation of $\alpha_1 - e_1 = R_{2\to 1} - R_{1\to 2} = e_2 - \alpha_2$, indicating that the broken balance of $\alpha_{1,2} \neq e_{1,2}$ could be realized in nonreciprocal systems when $R_{1\to 2} \neq R_{2\to 1}$. In practice, the symmetric incident angles $\pm\theta$ mimic two radiative channels, thereby nonreciprocal thermal radiation properties can be directly evaluated by the difference between $R_{-\theta}$ and $R_{+\theta}$. In this regard, several theoretical works[16–20] have been proposed. One of the pioneering studies was proposed using magneto-optical (MO) n-InAs photonic crystals[20]. The MO material, whose permittivity is described by an asymmetric tensor $\vec{\varepsilon}(\hat{z}\mathbf{B})$ showing as

$$\vec{\varepsilon} = \begin{pmatrix} \varepsilon & i\eta & 0 \\ -i\eta & \varepsilon & 0 \\ 0 & 0 & \varepsilon_z \end{pmatrix} \quad (2)$$

with $\eta \neq 0$ when an external magnetic field **B** is applied along the z-axis to break electromagnetic reciprocity. In that case, near-complete violation of Kirchhoff's law at a specific angle ($\theta$=61.28°) has been theoretically shown, at a price of a high **B**=3T because of weak MO response. To mitigate the stringent requirement of external stimulus, Zhao et al.[19] further added SiC gratings sitting on thick n-InAs and metal layers, where the excitation of high-$Q$ guided modes helps maintain a strong angle-dependent nonreciprocity despite reducing an order of magnitude of **B**. Nevertheless, such high-$Q$ properties are sensitive to material loss[21], and the employed optical parameters (i.e., damping rate $\Gamma \sim 1\times10^{11}$ rad/s) in numerical calculations may be too ideal to be attainable for natural materials. Later on, topological Weyl semimetals with intrinsic magnetism are also employed to break the balance of emissivity and absorptivity theoretically without any external stimulus[17,22,23]. Such amazing properties of these topological materials may be attainable at an extremely low temperature but have not been experimentally validated for practical energy devices at room temperature. Up to now, direct experimental measurements or practical room-temperature designs of actual physical structures that can violate Kirchhoff's law are still elusive. Although Hadad et al.[24] has designed spatiotemporally modulated antennas and detected asymmetric absorption and emission spectra in GHz, the paradigm to realize strong nonreciprocal thermal radiation at a higher optical-modulation frequency is still elusive and unclear. Notably, one critical issue is the large loss (small electron mobility $\mu$) in natural materials leading to broad resonances in spectra. In order to well separate the emission and absorption peaks in **Fig. 1b** (right), a strong external magnetic field must be required consequently. Thereby, how to realize large nonreciprocal radiation under a moderate external magnetic field in nature MO systems is of significant importance and value, but yet answered. On the other hand, existing nonreciprocal thermal emitters just show asymmetric spectra within a single and narrow band[17,19,20,22,23] (**Fig. 1b**). Shaping nonreciprocal radiation spectra with more diversified applications, e.g., multi-band and broadband nonreciprocal emission that would be more practical in several energy devices, remains an open challenge in both theoretical and experimental points of view.

To overcome these challenges, we turn our attention to magnetized epsilon-near-zero (ENZ) materials[25–33] which show advantages in strengthening both MO response and light-matter interaction within nanofilms. Since the diagonal elements (i.e., $\varepsilon$ in permittivity tensor $\vec{\varepsilon}$) crosses zero at a specific frequency, it enables to significantly enhance MO strength measured by Voigt parameter $\varsigma = \dfrac{\eta(\hat{z}\mathbf{B})}{\mathrm{Re}(\varepsilon)\to 0} \to \infty$ even in the presence of a small $\eta(\hat{z}\mathbf{B})$. In addition, an ultrathin ENZ film shows superior performance in localizing electromagnetic energy within a thin layer owing to the boundary condition $E_{ENZ} = \frac{E_i \varepsilon_i}{\varepsilon \to 0} \to \infty$ ($\varepsilon_i$ and $E_i$ are the permittivity and electric field of the adjacent layer), improving light absorption ability. Several works have explored the role of ENZ features in thermal radiation properties[31] in reciprocal systems, including but not limited to wavelength-selective thermal emitters[34], broadband directional emission[35] and so on. Interestingly, the abilities of ENZ behaviors combined with MO effects in shaping nonreciprocal thermal radiation have barely been investigated in both theory and experiment, preventing us from accessing many novel physics and potential applications.

Utilizing the magnetized and ultrathin ($< \lambda/40$, $\lambda$ is the working wavelength of interest) ENZ InAs semiconductor layers, here we report, for the first time, the experimental evidence that remarkable violation of Kirchhoff's law has been observed under a moderate external magnetic field. The evolution of nonreciprocal spectra induced by asymmetric Berreman modes (BMs) can be tailored by changing either doping level or thickness of thin InAs layers. Apart from single-band nonreciprocity (**Fig. 1b**) merely numerically reported before, we also demonstrate nonreciprocal emission/absorption spectra with dual-band (**Fig. 1c, left**) and broadband features (**Fig. 1c, right**) on the magnetized ENZ platform by utilizing dielectric gratings and gradient ENZ layers, respectively. We thus fabricated a series of photonic structures atop the ultrathin InAs films to experimentally realize the proposed versatile nonreciprocal emitters, whose spectral and angular responses are also characterized and compared with reciprocal counterparts.

As shown in **Fig. 1d**, we first consider an opaque and nonreciprocal thermal emitter composed of a thin InAs semiconductor film (thickness *t*) with a doping concentration of $n_{e,H} \approx 7.4 \times 10^{17}$ cm$^{-3}$ epitaxially grown atop the highly-doped InAs substrate ($n_{e,s} \approx 1.6 \times 10^{18}$ cm$^{-3}$, electron mobility $\mu_s = 7890$ cm$^2$/VS). The details of sample fabrication processes can be found in *Materials and Methods*. **Figure 1e** gives the permittivity of one fabricated InAs film calculated by the measured values of $n_{e,H}$ and $\mu \approx 9000$ cm$^2$/VS, exhibiting a zero crossing in Re($\varepsilon$) around $\omega_{ENZ,H} = 7.56 \times 10^{13}$ rad/s. The details of measured optical parameters and calculation methods of InAs materials are given in *Supplementary text S1*. Here, the highly-doped InAs substrate serves as a metal-like reflector within the waveband of interest and also ensures a better quality of deposited InAs films due to lattice matching mechanism. Such a thin ENZ film on a substrate is capable of supporting guided modes under TM polarization ($E_x, k_y, H_z$), whose dispersion relation is governed by

$$\tan(\kappa t) = \frac{\kappa \varepsilon (\gamma + \frac{\gamma_s}{\varepsilon_s})}{\varepsilon k_0^2 - \beta^2 - \eta \beta \cdot (\gamma - \frac{\gamma_s}{\varepsilon_s}) - (\varepsilon^2 - \eta^2)\gamma \cdot \frac{\gamma_s}{\varepsilon_s}} \quad (3)$$

where $\beta$ represents the propagation constant, $\kappa = \sqrt{k_0^2 \frac{\varepsilon^2 - \eta^2}{\varepsilon} - \beta^2}$ and $\gamma = \sqrt{\beta^2 - k_0^2}$ ($k_0 = \frac{\omega}{c} = \frac{2\pi}{\lambda}$, $c$ is the light speed in vacuum). The dispersion derivation can be found in *Supplementary text S2*. The $\varepsilon_s$ is permittivity of the substrate and $\gamma_s = \sqrt{\beta^2 - k_0^2 \varepsilon_s}$. **Figure 1f** compares dispersion curves of a 700 nm ($t < \lambda/40$) InAs film on PEC substrate ($\frac{\gamma_s}{\varepsilon_s} \to 0$) in the lossless limit ($\mu \to \infty$). The radiative BMs are located within the light cone, which can directly couple with free-space propagating modes and introduce a resonant peak in far-field absorption/emission spectra. The surface ENZ modes below the light line should be excited, i.e., with the assistance of dielectric gratings. In general, the dispersion curves are perfectly symmetric for reciprocal cases [$\omega(-\beta) = \omega(\beta)$, black lines]. In the presence of **B**, the dispersion profile of BMs within the light cone experiences a rotation with respect to the center of $(\omega, \beta) = (\omega_{ENZ}, 0)$ in **Fig. 1f**. The rotation center is exactly known as bound states in the continuum (BICs)[21,36] in the lossless limit, at which the root and pole of **Eq. (3)** are perfectly overlapped[37]. The extent of rotation increases with a higher **B**. The employed highly-doped InAs substrate cannot be seen as a perfect mirror thus will introduce a small angular shift of the rotation center as discussed in *Supplementary text S3*, which further contributes to observing large nonreciprocity in radiation spectra. As for a pair of symmetric wavevectors: $\beta(\pm\theta)$, the corresponding frequencies of modes are largely separated and move along opposite directions as the **B** increases. Such symmetry-broken modes also provide a new mechanism to shape nonreciprocal emission and absorption spectra as discussed below. Note that the magnitude of material loss in the top thin InAs film plays an important role in violating Kirchhoff's law. As mentioned in previous theoretical works[19], one was always searching for a high-$Q$ resonance so as to relieve the requirement of a high external magnetic field. In theory, the $Q$ factor of modes is determined by both radiative and non-radiative channels[38]: $Q^{-1} = Q_N^{-1} + Q_R^{-1}$, where a high $Q_N$ requires low-loss materials. A large loss in natural materials associated with broad resonant peaks will conceal the expected nonreciprocal phenomena, making it difficult to detect and observe experimentally. In our experiments, the damping rate $\Gamma$ of InAs films is about $5.9\times10^{12}$ rad/s. Although it is still an order of magnitude larger than the values taken in theoretical works[19,20], we still successfully realize large violation of Kirchhoff's law even under a moderate **B**, benefiting from synergizing MO effects and ENZ properties.

To explore the role of asymmetric BMs-induced nonreciprocity in thermal radiation, we fabricated samples with different thicknesses *t* at two doping levels: high $n_{e,H} \approx 7.4\times10^{17}$ cm$^{-3}$ and low $n_{e,L} \approx 6.5\times10^{17}$ cm$^{-3}$. Here, we define $\alpha(\theta) = 1 - R_{-\theta}$ and $e(\theta) = 1 - R_{\theta}$ in what follows, according to energy conservation. The reflectivity spectra were measured by a custom-built infrared MO Kerr effect characterization set-up in **Fig. 2a**. The details of experimental set-up are given in *Methods*. Using the measured parameters of InAs materials, **Fig. 2b** first calculates the angular emission and absorption spectra in the case of *t*=700 nm and **B**=1.5T, based on rigorous coupled-wave theory[39]. As expected, Kirchhoff's law of thermal radiation is violated, where radiation spectra have been re-shaped with enhanced emissivity but suppressed absorptivity spectrally and angularly. The typical radiation spectra for this case at $\theta = 60°$ (**Fig. 2c**) and angular pattern at $\omega_0$ (**Fig. 2d**) with broken symmetry are also presented. The weak resonant peaks on the left panel in **Fig. 2b** could be ascribed to the large material loss in InAs thin film. As revealed in our previous work[21], the presence of material loss in such a system will make the BICs (red triangle in **Fig. 1f**) split into two topological singularities in reflection phase diagram, which exactly correspond to perfect absorption/emission points. Increasing the material loss makes the phase singularity move away from the BIC origin along the dispersion curves, consequently moving out of light cone for a large material loss and resulting in a significant reduction of absorptivity in **Fig. 2b** (see detailed analysis in *Supplemental text S4*). Further, the measured data for samples with different doping concentrations and thicknesses ($t$=200, 700, 1000 nm) at $\theta = 60°$ are compared in **Figs. 2e** ($n_{e,H}$) and **2f** ($n_{e,L}$). The nonreciprocal BMs-induced resonant peaks red-shift to the low-frequency band in **Fig. 2f** since the low-$n_{e,L}$ condition makes the ENZ frequency moves to $\omega_{ENZ,L} = 7.1 \times 10^{13}$ rad/s. The grey lines show the overlapped emission and absorption in the absence of **B**. For the measured cases, a larger thickness ensures a higher value of $|e(60°) - \alpha(60°)|_{max}$, whose maximum can be up to 0.63 at $t$=1000 nm. The shadowed light green areas denote the waveband of $|e(60°) - \alpha(60°)| \geq 0.3$, showing superior nonreciprocal radiation performance within the waveband of interest. In reciprocal thin ENZ structures, there is an optimal angle (or thickness) where the perfect absorption can be available at a fixed thickness (angle). Such behaviors can also be understood from the perspective of topological phase singularities (perfect absorption points), whose spectral and angular positions are also influenced by the radiative loss of the system that could be modified by the thickness $t$ and angle $\theta$. Similar features can also be observed in nonreciprocal systems as presented in **Fig. 2g**, where the results of $|e(\theta) - \alpha(\theta)|_{max}$ are given as the functions of **B** and thickness $t$ at $\theta = 60°, 45°$. The measured data agree well with the simulated results. The measured asymmetric radiation spectra of $n_{e,H}$-samples with different thickness at $\theta = 45°$ are also given in **Fig. S12** for comparison. Our realized BM-induced nonreciprocal emission can be preserved over a wide angle. Besides, we also observe apparently nonreciprocal resonant peaks with $|e(60°) - \alpha(60°)|_{max} \approx 0.45$ induced by BMs in highly-doped InAs substrate in **Fig. S8**, whose ENZ frequency is high ($\omega_{ENZ,s} = 11.2 \times 10^{13}$ rad/s).

As revealed in **Fig. 1f**, the asymmetric surface ENZ modes below the light line also show great potential in shaping nonreciprocal spectra. Here, we consider adding dielectric gratings sitting on the above ultrathin InAs films ($t$=700 nm) as depicted in **Fig. 3a**, so as to observe two separated resonant bands in emission and absorption spectra. **Figure 3b** shows the SEM image of fabricated aSi gratings with $p_g$=9 μm, $t_g$=1.05 μm, $t_s$=0.25 μm and $w_g$=0.5$p_g$, whose simulated asymmetric angle-resolved spectra at **B**=1.5T are shown in **Fig. 3c**. Apart from the BMs-induced asymmetric resonance at a low-frequency range, there are other two emission and absorption peaks owing to asymmetrically folded bands from the dispersion of surface ENZ modes in **Fig. 1f**. The surface-ENZ-modes-induced nonreciprocal properties become sensitive to the grating's period (**Fig. S11**). The separated peaks gradually red-shift as the period increases, but the locations of BM-induced nonreciprocal peaks remain robust. Such behaviors ensure us to shape two resonant mechanisms individually. The symmetric radiation spectra are also given in **Fig. S14** for better comparison. Although the addition of gratings looks similar to what was proposed by Zhao *et al.*[19], the origins of guide modes become distinct. The thickness of InAs interlayer in Ref. [19] is on the order of working wavelength for enhancing MO effects. In contrast, the ultrathin InAs film ($t < \lambda/40$) proposed here ensures the co-existence of two types of nonreciprocal resonances near the ENZ frequency. Note that the periodicity-sensitive nonreciprocal peaks are also susceptible to material loss. In our experiments, the optical index of deposited aSi is about 3+0.5i, leading to a wide overlap between $\alpha(60°)$ and $e(60°)$ around $\omega$=8.5×10$^{17}$ rad/s in **Fig. 3d**. The BMs-induced nonreciprocity is still strong with $|e(60°) - \alpha(60°)|_{max} \approx 0.45$. There is still plenty of room for achieving narrower peaks when lower-loss dielectric materials and more optimized designs could be found and used. Besides, more nonreciprocal resonant peaks can also be available by designing different grating patterns, i.e. asymmetric gratings, 2D gratings, meta-units [3], *etc.*

The proposed ultrathin magnetized ENZ film shows outstanding merits in shaping broadband nonreciprocal radiation properties. We showcase the gradient magnetized ENZ films composed of multilayer InAs films in **Fig. 4a**. The thickness of each layer is set as $t_i=0.35\mu m$, and the doping concentration in different layers gradually decreases from the bottom to the top. **Fig. 4b** shows the real part of diagonal element $\varepsilon$ for different doping concentrations, where the gradient $n_e$ leads to a shift of ENZ frequency. The grey shadowed region denotes the condition of $|\text{Re}(\varepsilon)|<1$, so that one can expect to break Kirchhoff's law over a wide spectral range thanks to the co-operative BMs in each thin InAs film. In our experiments, we fabricated two samples with three InAs layers with gradient $n_e$: $n_{e1} \to n_{e2} \to n_{e3}$ (**Figs. 4c,d**) and $n_{e2} \to n_{e3} \to n_{e4}$ (**Fig. 4e**), where $n_{e1}=8\times10^{17}$ cm$^{-3}$, $n_{e2}=7.5\times10^{17}$ cm$^{-3}$, $n_{e3}=7\times10^{17}$ cm$^{-3}$ and $n_{e1}=6.5\times10^{17}$ cm$^{-3}$. The simulated emission and absorption spectra are compared in **Fig. 4c.** The broadband emission over a wide angular range is presented, covering from the whole waveband of interest, along with totally suppressed absorption in both spectral and angular aspects. The measured broadband nonreciprocal spectra at $\theta=60°$ are shown in **Fig. 4d** (left), where the difference of $|e(60°)-\alpha(60°)|\geq 0.3$ can be maintained over nearly a bandwidth of 2.83 μm. Similar behaviors can be available for the other case of $n_{e2} \to n_{e3} \to n_{e4}$ with $\Delta\lambda \approx 2.73\mu m$ (**Fig. 4d**, right). The corresponding dispersion profiles are analyzed in **Fig. S9**. The width of nonreciprocal spectra can also be further broadened when considering more gradient-$n_e$ layers. In **Fig. 4e**, we evaluate the $\Delta\lambda$ of $|e(60°)-\alpha(60^0)|\geq 0.3$ for structures with different numbers of magnetized ENZ films. For example, the 10-layer structure contains gradient doping concentrations from $4\times10^{17}$ cm$^{-3}$ (top) to $8.5\times10^{17}$

cm$^{-3}$ (bottom) with an interval of $0.5\times10^{17}$ cm$^{-3}$. In this case, the broadband nonreciprocal emission and absorption can be maintained covering $\Delta\lambda\approx12.8\,\mu\text{m}$ (inset). The spectral position can also be tailored using different combinations of $n_e$ (**Fig. S10**). The proposed strategies for violating Kirchhoff's law covering a wide band could be immediately applied to several applications, such as radiative cooling, thermal illusion, thermal camouflage, etc.

To further evaluate the polarization-sensitive behaviors of proposed nonreciprocal thermal emitters, here we also measured and compared the emission and absorption spectra of fabricated samples under TE polarization ($H_x, k_y, E_z$). As expected, as for single-layer and multilayer magnetized ENZ structures, the obtained emissivity and absorptivity are consistently equal and extremely low as shown in **Fig. S16** since there is no MO response for TE waves. Large violation of Kirchhoff's can still be distinct for polarization-averaged emission and absorption spectra. This capability further guarantees the direct thermal emission of proposed thermal emitters with desirable nonreciprocal behaviors.

We have provided experimental evidence that a large violation of Kirchhoff's law can be realized under a moderate magnetic field using ultrathin ENZ films ($<\lambda/40$). Our utilization of magnetized ENZ structures also provides a versatile platform to shape nonreciprocal emission and absorption spectra in a more diverse manner: the co-excitation of asymmetric BMs and surface ENZ modes ensures the realization of dual-band nonreciprocal spectra, and broadband violation of Kirchhoff's law has also been experimentally realized using magnetized gradient ENZ multilayers. We emphasize that the working wavelength of proposed nanostructures can be tailored by flexibly modifying the doping level of employed materials, including but not limited to other III-IV semiconductors like GaAs, InP, and so on. In addition, the proposed concepts and revealed mechanism herein for tailoring nonreciprocal thermal radiation can also be extended to other metamaterials and metasurfaces combined with MO effects, like hyperbolic metamaterials[40], zero-index metamaterials[28,33,41–44], epsilon-near-pole materials[45]. Besides, it is also believed that the proposed magnetized ENZ platforms integrated with Weyl semimetals[46] may be possible to break Kirchhoff's law without any external stimulus, showing potentials in designing compact and nonreciprocal micro/nanodevices. More importantly, our results lay a solid experimental foundation for more comprehensive and general studies in nonreciprocal thermal radiation control, which will breed new possibilities in renovating several advanced energy devices like solar cells [15,47], thermophotovoltaic [12,48–50], radiative cooling[7–9], etc.

**Materials and Methods**

**Photonic structure fabrication**

The InAs materials were grown using a VG Semicon V90 gas source molecular beam epitaxy (GSMBE) system. A VG thermo-cell with two heat zones was used for In sources to reduce surface oval defects. The group V source was obtained by cracking Arsine ($AsH_3$) to $As_2$ at 1000 °C. Elemental silicon was used as a n-type dopant. The 2-inch n-type substrates used here were S-doped (100) InAs with a carrier concentration of ~$1.6 \times 10^{18}$ cm$^{-3}$ and a thickness of 500 μm. Before growth, a heat treatment at 365 °C for 3 minutes under the molecular beam of $As_2$ was carried out to remove the surface oxide. Then the InAs epitaxial layers were grown on InAs substrates at a growth temperature of 360 °C. The InAs growth rate and doping were controlled by changing the temperature of the In cell and the temperature of the Si cell, respectively. In our experiments, the growth rate for InAs layers was kept at 1 μm/h. The epitaxial layer thicknesses for samples discussed in Fig. 2 are 200, 700, and 1000 nm, respectively, with expected doping concentrations of $8 \times 10^{17}$ cm$^{-3}$ and $7 \times 10^{17}$ cm$^{-3}$. The doping levels of samples were calibrated by Hall measurements for InAs samples grown on semi-insulating InP substrates. Note that, in the main text, we use the fitting value of $n_{e,H} \approx 7.4 \times 10^{17}$ cm$^{-3}$ (Fig. 2e) and $n_{e,L} \approx 6.5 \times 10^{17}$ cm$^{-3}$ (Fig. 2f). As for gradient structures with three InAs layers, the thickness of each layer is 350 nm. From the bottom to the top layer, the targeted doping concentrations are gradually decreased from 8, to 7.5 and finally $7 \times 10^{17}$ cm$^{-3}$ for the samples in Fig. 4d. The amorphous silicon layer was deposited via magnetron sputtering. The power was set to 90W and the gas pressure was $5\times10^{-3}$ mbar controlled by argon. The grating shapes were fabricated on silicon layer by photolithography. Using the photoresist as a mask, the silicon grating was obtained via deep reactive ion etching. Finally, the remaining photoresist was removed by acetone.

**Measurement Setup**

The reflectivity spectrum was measured in a custom-built infrared magneto-optical Kerr effect characterization set-up, as shown in Fig. 2a. Broadband infrared light with the wavelength range of 10 μm to 30 μm was generated from a Fourier transform infrared spectroscopy (FTIR, Bruker TENSOR27) and incident into the free space. A gold coated mid-infrared enhanced plane mirror (PF20-03-M02, THORLABS) was used to reflect the light by 90°. Then a 90° off-axis parabolic mirror coated with silver (MPD249H-P01, THORLABS) was used to deflect the beam. A pair of 30° off-axis parabolic mirrors with focal length of 10 inches (MPD2103-P01, THORLABS) were used to focus the beam on the sample surface and to collect the reflected light from the sample, respectively. Finally, another 90° off-axis parabolic mirror was utilized to focus the light onto the detector. Static magnetic field was applied parallel to the sample surface and perpendicular to the incident plane using a custom-built electromagnet with maximum applied magnetic field up to 2T. For background calibration, the reflectivity of a silicon wafer coated with 200 nm thick gold film was first measured on the set-up and considered as unity. The samples' reflectivity spectra were normalized to this background. Each reflection spectrum in this paper was presented by the average value of three consecutive measurements on the same sample. The error bars were the standard deviation of these measurements.

**Acknowledgment**

C.-W. Q. acknowledges the financial support of the Ministry of Education, Republic of Singapore (No. R-263-000-E19-114). C. Y. Z. acknowledges the National Natural Science Foundation of China (No. 52120105009 and No. 51906144) and the Shanghai Key Fundamental Research Grant (No. 20JC1414800). M. Q. L.

acknowledges the support from the SJTU-NUS Joint PhD Project. L. B., S. X. and J. Q. acknowledge the financial support by Ministry of Science and Technology of the People's Republic of China (No. 2018YFE0109200), the National Natural Science Foundation of China (Nos. 51972044, 52021001 and 52102357), Sichuan Provincial Science and Technology Department (Nos. 2019YFH0154 and 2020ZYD015), and the Fundamental Research Funds for the Central Universities (No. ZYGX2020J005). H. L. acknowledges National Natural Science Foundation of China (Nos. 61875220 and 62022084), Chinese Academy of Sciences (Nos. ZDBS-LY-JSC009 and YJKYYQ20200032), and Science and Technology Commission of Shanghai Municipality (No. 20XD1424700).

**Author Contributions**

M. Q. L., C. Y. Z., and C-W. Q. conceived the ideas. M. Q. L. performed the simulations and advised in the experimental design. L.B. and H.L. led the experiments. W.J.W. fabricated the samples. S. X. performed the optical measurements. M.Q.L., S.X., J.Q. H. L., L. B., and C-W. Q analyzed the data and all authors discussed the results. M. Q. L. wrote the manuscript with inputs and comments from all authors. H. L., L. B., C. Y. Z., and C-W. Q. supervised the project.

**Competing interests:** The authors declare no competing interests.

**Data and materials availability:** All associated data and materials are available in the manuscript and supplementary materials.

**Supplementary Materials**

S1. Optical properties calculations of fabricated samples

S2. Dispersion derivation of magneto-optical nanostructure

S3. Impact of the substrate on re-shaping asymmetric dispersion

S4. Impact of material loss in ENZ films on nonreciprocal thermal radiation

S5. Nonreciprocity induced by high-doped InAs substrate around ENZ waveband

S6. Broadband nonreciprocal thermal emission in gradient ENZ layers

Fig. S1 to S16

Figures:

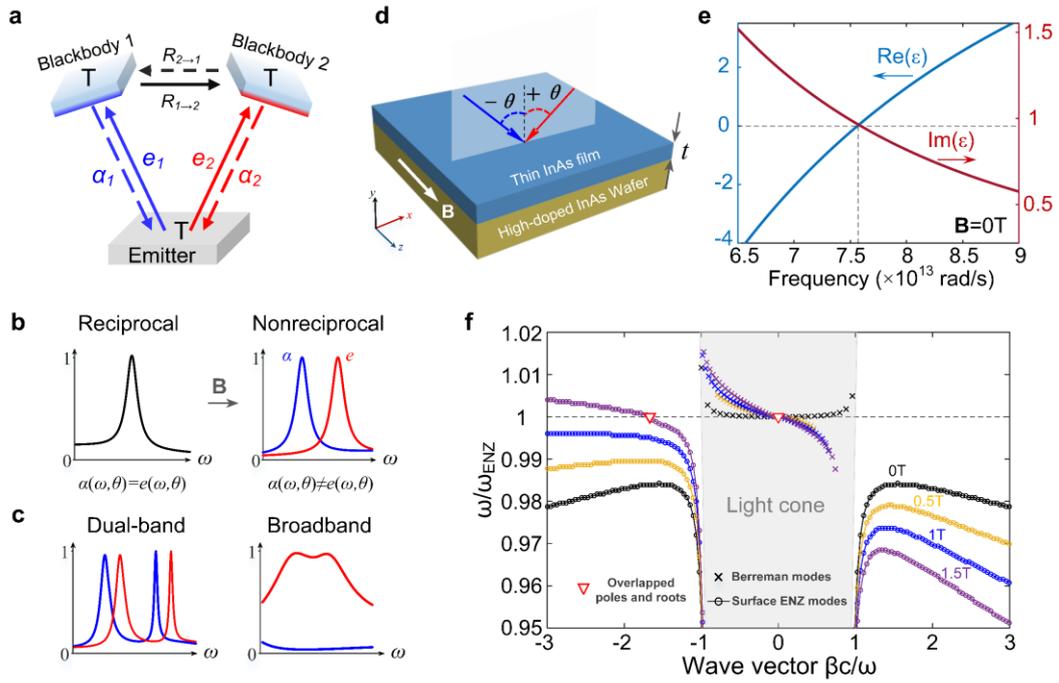

**Fig. 1 Schematics of nonreciprocal thermal radiation control. a**, Energy flow diagram of a general thermal emitter. **b**, Compared radiation spectra of reciprocal (left) and nonreciprocal (right) thermal emitters. The overlapped spectra are separated in the presence of **B**. **c**, Proposed nonreciprocal thermal emitters with dual-band (left) and broadband properties (right). **d**, A schematic of a single ultrathin magnetized ENZ InAs film (thickness $t$) atop the reflective high-doped InAs substrate. The external magnetic field **B** is applied along the $z$-direction. The $\theta$ is incident angle in experiments. **e**, The dielectric permittivity of fabricated InAs film with $n_{e,H}$ and $\mu$ in the absence of **B**. **f**, Dispersion relation of a magnetized ENZ film ($t$=700 nm) on PEC substrate in the lossless limit, under different values of **B**: black (0T), yellow (0.5T), blue (1T) and purple (1.5T).

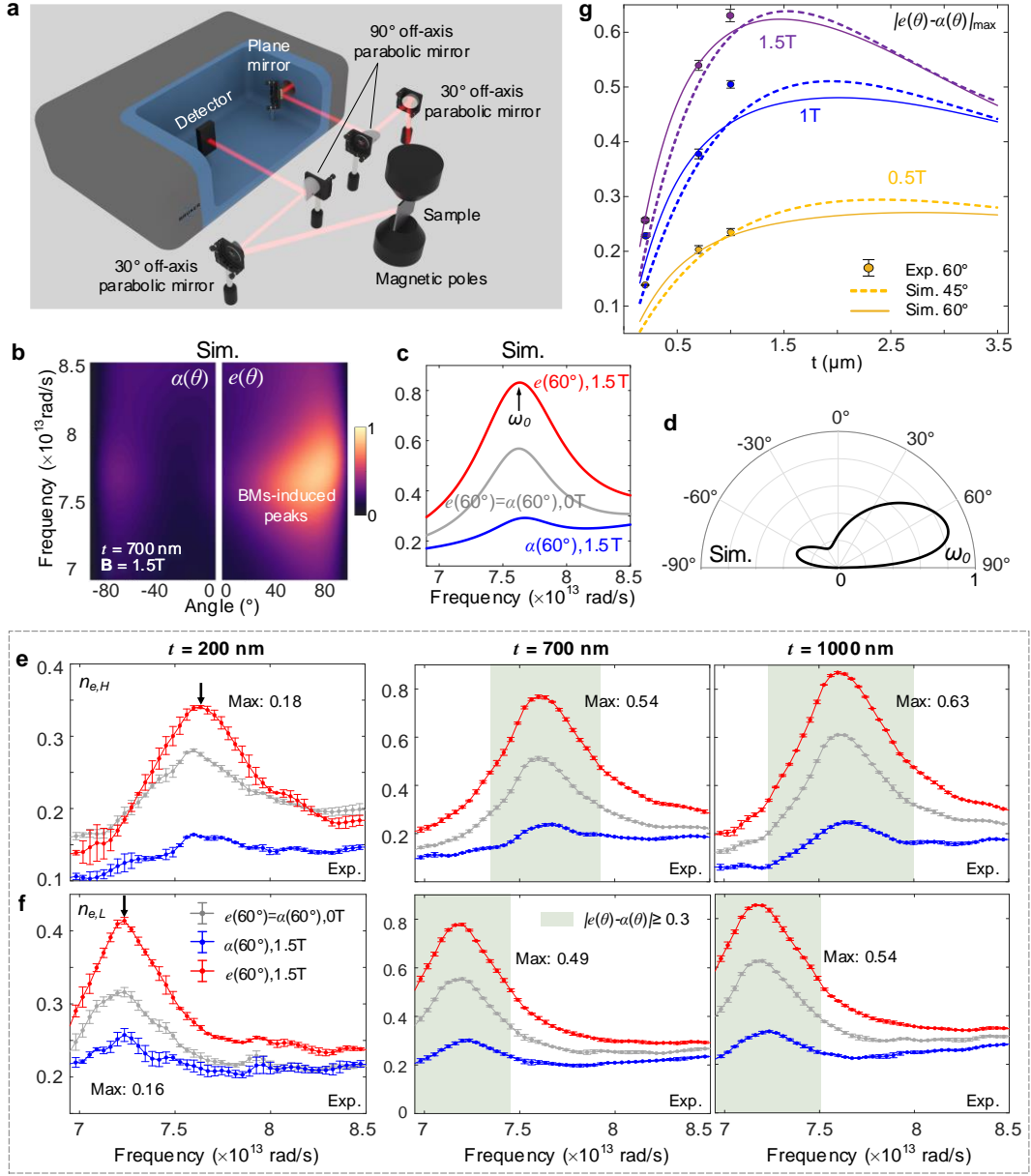

**Fig. 2 Breaking Kirchhoff's law induced by asymmetric Berreman modes. a**. The custom-built infrared MO Kerr effect characterization set-up. **b**. Simulated nonreciprocal absorption (left) and emission (right) spectra in the case of $t$=700nm, **B**=1.5T and $n_{e,H}$, along with **c**, asymmetric spectra at $\theta=60°$ and **d**, angular pattern at $\omega_0$. Measured nonreciprocal emission (red) and absorption (blue) spectra for samples with **e**, a high $n_{e,H}$ and **f**, a low $n_{e,L}$ at different thicknesses: $t$=200 nm (left), $t$=700 nm (middle) and $t$=1000 nm (right). The grey lines present the overlapped emission and absorption spectra without **B**. The maximum difference of $|e(60°)-\alpha(60°)|$ is marked with black arrows or values. **g**, The simulated and measured

results of $|e(\theta)-\alpha(\theta)|_{max}$ as a function of thickness $t$, working angles (full lines for $\theta = 60°$, dashed lines for $\theta = 45°$), and the magnitude of **B**, for the high-$n_{e,H}$ samples.

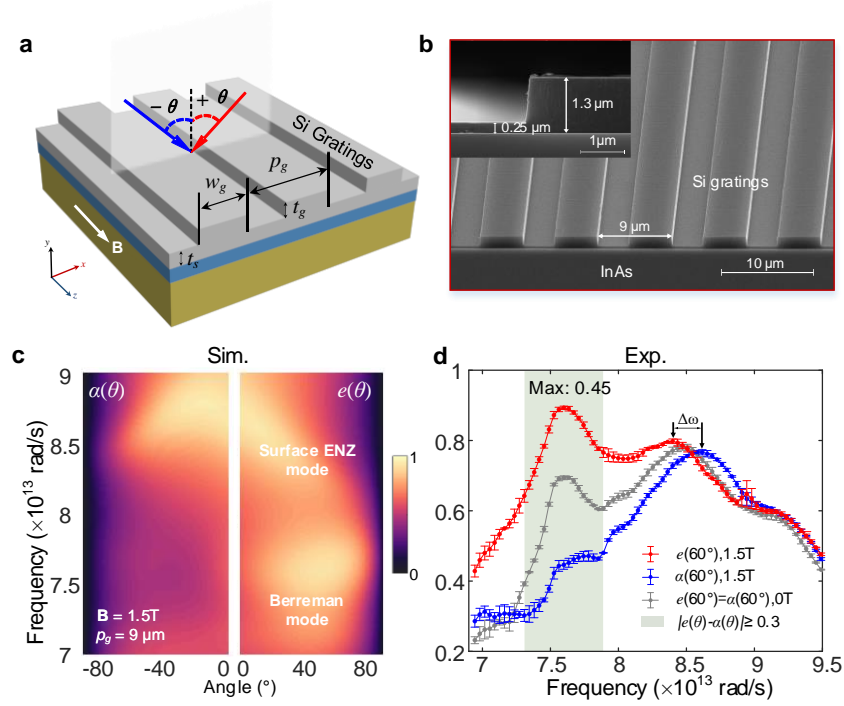

**Fig. 3 Dual-band nonreciprocal thermal emitters. a**, A schematic of dielectric gratings on the previous single-layer structure in Fig. 1d ($t$=700 nm). The thickness of the dielectric interlayer is $t_s$. The period, height and width of gratings are $p_g$, $t_g$ and $w_g = 0.5 p_g$, respectively. **b**, The SEM images of fabricated gratings at $t_s$ =0.25 μm, $t_g$ =1.05 μm, $p_g$ =9 μm. **c**, Simulated angle-resolved nonreciprocal absorption (left) and emission (right) spectra of the sample in **b**, along with **d**, the measured nonreciprocal spectra at $\theta = 60°$ and **B**=1.5T. The grey line presents the overlapped emission and absorption spectra without **B**. The peak shift induced by asymmetric surface ENZ modes is about $\Delta\omega = 0.16\times 10^{13}$ rad/s.

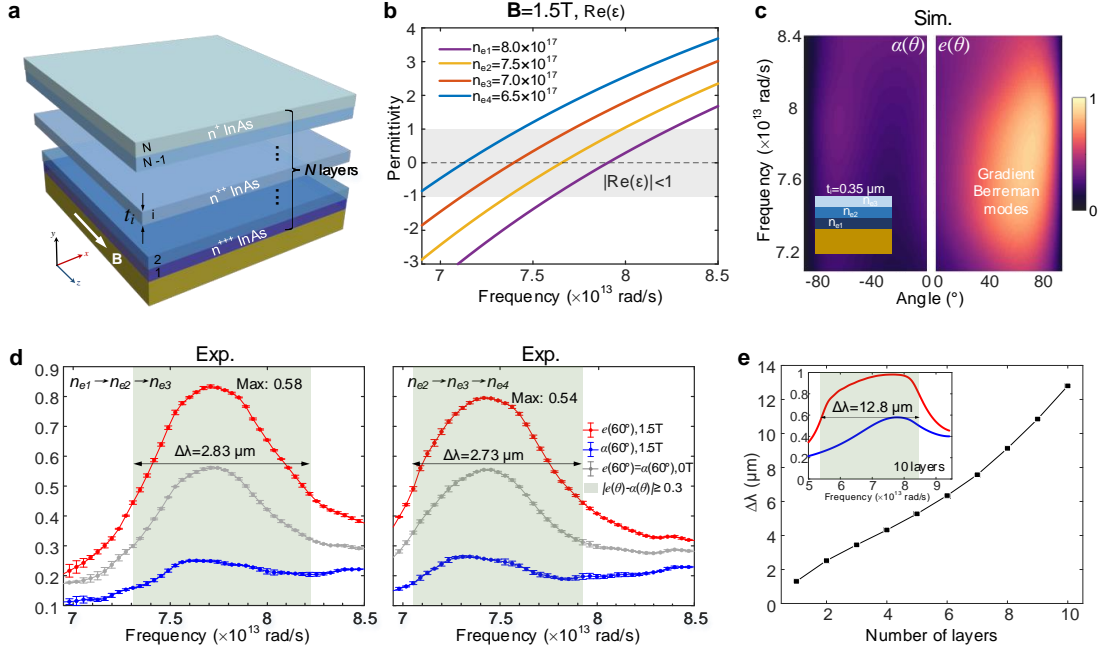

**Fig. 4 Broadband nonreciprocal thermal emitters. a**, A schematic of multilayer magnetized ENZ structures with gradient $n_e$. The value of doping concentration gradually decreases from the bottom layer. The thickness of layer is denoted as $t_i$. **b**, The real part of doped InAs film with different doping concentrations $n_e$. The ENZ region is shadowed in grey ($|\text{Re}(\varepsilon)|<1$). **c**, Simulated angle-resolved nonreciprocal absorption (left) and emission (right) spectra of a 3-layer structure with $t_i$=0.35μm gradient $n_e$ from the bottom to the top layers: $n_{e1} \to n_{e2} \to n_{e3}$. **d**, Measured nonreciprocal emission (red) and absorption (blue) for sample 1 (left) and sample 2 (right). The grey lines present the overlapped emission and absorption spectra without **B**. The light blue shadowed regions denote the waveband with $|e(60°)-\alpha(60°)|\geq 0.3$. **e**, Calculated results of $\Delta\lambda$ for $|e(60°)-\alpha(60°)|\geq 0.3$ as a function of the number of layers ($t_i$=0.35μm) with gradient $n_e$. In the case of 10-layer structures, the range of $n_e$ is from $4\times10^{17}$ cm$^{-3}$ (top) to $8.5\times10^{17}$ cm$^{-3}$ (bottom) with a fixed interval $0.5\times10^{17}$ cm$^{-3}$ between adjacent layers. The $N$-layer structure is obtained by removing the top 10-$N$ layers.